\documentstyle[12pt]{article}
\begin{document}
\baselineskip=21pt
\rightline{YCTP-P32-92}
\rightline{Aug 4 1992}
\vskip .1in
\begin{center}
{\large{\bf Angular Diameters as a Probe of a
Cosmological Constant and $\Omega$}}
\end{center}
\vskip .1in
\begin{center}
Lawrence M.
Krauss\footnote{Research supported in part
by the NSF,DOE, and TNRLC Bitnet:Krauss@yalehep}

{\it Center for Theoretical Physics and Department of Astronomy, Sloane
Laboratory}

{\it  Yale University, New Haven, CT 06511}

and

David N. Schramm\footnote{Research supported in part
by the NSF,DOE, and NASA}

{\it Departments of Physics, Astrophysics, and Enrico Fermi Institute}

{\it  University of Chicago, Chicago IL}

\end{center}

\vskip .1in

\centerline{ {\bf Abstract} }

\baselineskip=18pt

\noindent

The lensing effect of curved space, which can cause the angular
diameter of a fixed reference length seen on the sky to reach a minimum
and then increase with redshift, depends
sensitively on the value of the cosmological constant, $\Lambda$, in a flat
universe.  The redshift of an observed minimum and the asymptotic slope
can in principle provide strong constraints on $\Lambda$. The
sensitivity to a non-zero cosmological constant in a flat universe is compared
t
   o
the sensitivity to $q_0$ in an open universe without a cosmological constant,
an
   d to
inherent ambiguities due to uncertainties in distance measures and the possible
effects of evolution.  If evolutionary uncertainties can be overcome, the
reported observations of the angular diameter of compact radio jets as a
functio
   n of
redshift, which appear to exhibit such a minimum, could provide the strongest
available limit on the cosmological constant in a flat universe, and on
$\Omega$
in an open universe.

\baselineskip=21pt
\newpage

The Cosmological Constant may be at the same time the most strongly constrained
and the poorest understood theoretical quantity in nature.  The fact that we do
not live in an identifiably de Sitter universe implies that the Cosmological
Constant is over 120 orders of magnitude smaller than the value one might
naively expect, namely $\Lambda \approx M_{Pl}^4$ (using units where $h/2\pi
=c=1$). It is therefore tempting to speculate that its value is identically
zero
   .
However, not only do we have no clear understanding of why this might be the
cas
   e
(although recently some progress on this issue has been made [Baum 1983;Hawking
1984;
 Coleman 1984]), but there are
observational reasons which suggest that the Cosmological constant might
 {\it not} be zero, but could dominate the energy density of the universe
today.  If the oldest globular cluster stars are indeed older than
15 Gyr, as current analyses suggest, this will be incompatible with any value
of Hubble constant greater than about 40 km/sec/Mpc in a flat universe without
a
cosmological constant, or about 70 km/sec/Mpc in an open universe.  Similarly,
structure formation arguments in a Cold Dark Matter dominated universe with
scal
   e
invariant adiabatic density perturbations, normalized to the recent COBE report
   of a
quadrupole anisotropy [Smoot et al 1992], also improve if there exists a
non-zer
   o
cosmological constant.

Recently, a number of researchers have examined the possibility of using the
optical depth for gravitational lensing of distant quasars by intervening
galaxi
   es
to probe the geometry of the universe, and hence constrain the cosmological
constant [Turner 1990; Fukugita, Futamase and Kasai 1990; Mao 1991;
Krauss and White
1992; Fukugita and Turner 1992; Kochanek 1992].  The existing
data, which is sparse, appears to constrain a cosmological constant
contribution
    to
$\Omega$ to be less than about .95, but remains compatible with the favored
valu
   e
of $\Omega_{\Lambda} \approx 0.8$ [i.e. see Krauss 1992].  The redshift
distribu
   tion
of galactic lenses is also a useful probe [Fukugita, Futamase and Kasai 1990;
Krauss and White 1992; Kochanek 1992], but
here again, the available statistics are marginal.

We propose here to use another sort of gravitational lensing phenomenon for
this purpose.  It is well known that, due to its spatial curvature, the
universe
itself can act as a lens of large focal length.  Nearby objects are not
affected
   ,
but objects located at distances which approach the Hubble size can be greatly
magnified (i.e. see [Misner, Thorne, and Wheeler 1973]).  Traditionally
this effect has been discussed as a possible way to distinguish between an open
and closed universe.  We describe here how it can be used in principle to
effectively limit the cosmological constant, assuming the universe is
flat, as both theoretical prejudice and several recent analyses suggest.

If we write the Robertson-Walker metric in the form:
\begin{equation}
  ds^2 = dt^2 -
  R^2(t)\left(d\chi^2 + s_k^2(\chi) d\Omega^2 \right)
\label{eqn:metric}
\end{equation}
where the form of $s_k$ depends on the curvature ($k=0,\pm1$), the angular
diameter $\delta \ll1$ of a proper distance D located perpendicular to the line
   of
sight at $\chi =\chi_1, t=t_1$ is given by [Weinberg 1972, Misner Thorne and
Wheeler 1973]
\begin{equation}
  \delta = {D\over{R(t_1)s_k(\chi_1)}}.
\label{angl}
\end{equation}

In a flat $k=0, \Omega=1$ universe with non-zero cosmological constant
contribution $\Lambda = 1-\Omega_0$, $s_k(\chi)=\chi$, and the relationship
betw
   een
the co-ordinate $\chi$ and the redshift $z$ is given by
\begin{equation}
\chi \rm (\mit z\rm )=\int_{1}^{1+\mit z}{dy \over \rm ({\mit
\Omega }_{\rm 0}{\mit y}^{\rm 3}-{\mit \Omega }_{\rm 0}+1{)}^{1/2}}
\label{chiz}
\end{equation}

This integral cannot be written in a simple analytical form in general, except
f
   or
the extreme cases $\Omega_0 =0,1$.  In these cases, normalizing
$D/R(t_{today})=
   1$,
one finds:
\begin{equation}
\delta =  \left\{ \begin{array}{cc}{(1+z)^{3/2}\over2[(1+z)^{1/2} -1]} ;
&  \Omega_0 =1  \\  \\
{(1+z)\over z} ; &  \Omega_0 =0
\end{array} \right.
\label{anal}
\end{equation}

As can be seen from (\ref{anal}), in the standard case of a flat universe when
t
   he
cosmological constant is zero ($\Omega_0 =1$) the angular diameter
as a function of redshift has its
well known minimum at $z=5/4$ .  However when there is
zero matter ($\Omega_0=0$) and the cosmological constant contributes all the
ene
   rgy
density the minimum disappears.   For intermediate cases, the integral
(\ref{chi
   z})
must be done numerically.  Several such cases, along with the two extremes, are
shown in figure 1, plotted on a log-log plot.  As can be seen there, as long as
$\Omega_0 \ne 0$ the minimum persists, although it gets less pronounced, and is
pushed to higher redshifts as the matter density is decreased.

These features can be exploited in principle to constrain the cosmological
const
   ant,
as can be seen explicitly by differentiating the angular diameter versus
redshift curves.  The differentiated curves are shown in figure 2.  Here a
reference line representing zero slope is displayed.  As expected from the
analytic results for the extreme cases (using our normalization for $D$), the
asymptotic slope approaches $ 1/2 \ (0)$ for $\Omega_0 =1 \ (0)$
respectively. (In the former case, the slope actually overshoots its asymptotic
value in the region $ z<5$). The redshift of the minimum angular diameter (zero
slope) moves from $z =1.25$ for $\Omega_0 =1$ to $z \approx 2$ for $\Omega_0
=0.
   1$,
while the asymptotic slope moves from $\approx 1/2$ to $ \approx 0.1$ in this
case.  The specific redshifts for the minimum angular diameter, and the slope
of
the angular diameter-redshift function at z=4 are tabulated for various values
o
   f
$\Omega_0$ in table 1.

Of course, this test is not an unambiguous probe of $\Lambda$, unless we
also know the geometry of the universe.  For example, it is well known that
if the universe is open, the angular diameter can also continue to
decrease, although it is less well known that this need not be the case.  To
exa
   mine
the sensitivity to $\Lambda$ for a flat universe versus $\Omega$ for an open
universe, we can repeat the above analysis for a $k=-1$ cosmology, with zero
cosmological constant.  In this case $s_k(\chi) = sinh(\chi)$.  The formula for
$\chi$ as a function of redshift  is in this case (where $q_0$ is the
decelerati
   on
parameter =$\Omega_0/2$ for $\Lambda =0$):

\begin{equation}
\chi \rm (\mit z\rm )=[1-2q_0]^{1/2}\int_{1}^{1+\mit z}{dy \over y[2q_0y
+(1-2q_0)]^{1/2} }
\label{chiz2}
\end{equation}
which can be done analytically.  In the limit $q_0 =0$, one finds, using
(\ref{angl}) that the angular diameter decreases monotonically as a function of
reshift.  However, just as for the flat case with small non-zero matter
density,
    if
$q_0\ne0$ then one finds that this function has a minimum, which again becomes
l
   ess
pronounced and occurs at higher redshift as $q_0$ is decreased.  In figure 1
the
results for the case $q_0=0$ and $q_0=0.1$ ($\Omega \approx 0.2$) are shown.
In
figure 2 the differentiated curves are displayed for these two cases.   As can
b
   e
seen, the $q_0=0$ case is almost identical to the $\Lambda =1$ case, which is
no
   t
surprising.  The case of $q_0=0.1$ is very similar to the $\Lambda =0.9$ case.
As a result, if an astrophysical source with fiducial proper length can be
measu
   red
as a function of redshift out to redshifts of 2-4,{\it and if the possible
evolutionary uncertainties discussed below can be removed} a sensitivity which
c
   an
distinguish an $\Omega =1$ flat universe from an $\Omega=0.2$ open universe is
equivalent to a sensitivity which can rule out a value of $\Lambda >0.9$, if
one
assumes that the universe is flat.  If no other cosmological information were
utilized, one could not distinguish between these two possibilities using this
t
   est
alone.

It is worth stressing that a sensitivity to smaller values of $\Lambda$ is
not unreasonable.  The minimum of the angular diameter-redshift relation moves
from $z \approx 1.2$ to $z \approx 1.6$ for $\Lambda=0.7$.  Such a distinction
may prove plausible observationally.

These arguments would be primarily academic if there were not a realistic
possible
astrophysical fiducial probe.  Recently a survey of the angular
diameter of $\approx 100 $ compact parsec-scale radio jets (i.e. [Pearson
1990]) with size $O(40)$ pc ($ H=50$ km/s/Mpc) in AGN's using VLBI, extending
out to
 $z \approx 3$ has been reported [Kellerman 1992], in which a minimum at $z
\app
   rox 1$ is
apparently observed . As the parameters of this survey become
established, the analysis described here could be used to provide in principle
a
robust limit on the cosmological constant in a flat universe which might
plausibly exceed those presently available from the statistics of gravitational
lensing of quasars.  This of course depends in practice on how strongly such
effects as evolution can be constrained.

Indeed, any mention of a potential candidate probe of the geometry of the
univer
   se
would be incomplete without some discussion of the plagues on all such
measurements: evolutionary effects, and uncertainty in distance measures.
First
   ,
let us assume a minimum is observed. How can one be certain that the angular
diameter-redshift relation that one observes is geometric, and not related to
dynamical evolution of the source?  While no evolutionary relationship would be
expected to produce a minimum by itself, as we describe below, a  specific, but
plausible evolution of the size of compact jets could, when combined with the
geometric effect, efficiently mask, or mimic the behavior of angular
diameter vs redshift for all the cases we have examined.

Clearly, if the intrinsic size of compact jets varies monotonically with
redshif
   t,
this could shift the redshift of any observed minimum.  Unless this
variation is relatively smooth however, the slope of the angular
diameter-redshi
   ft
away from the minimum might be expected to be altered due to such dynamical
evolution in a distinguishable way.  The simplest, and at the same time
a physically plausible, possibility for an evolutionary trend which might mimic
   the
geometric effects is to imagine that the intrinsic source size D varies with
redshift in a way similar to the co-ordinate dependence, i.e. $D \approx D_0
(1+z)^{\alpha}$, where $D_0$ is the source size in nearby objects.

We find that such a variation can either produce or remove any pre-existing
minima, even for relatively mild values of $\alpha$.  To see that this is
possible, consider a $\Lambda=1$ flat universe, in which the slope of
$\delta(z)
   $
approached zero from below at large redshift.  If $\alpha=1/2$ in the relation
f
   or
$D$ above, then this slope would approach instead $1/2$ at large $z$, thus
mimicking the behavior of an $\Omega_0 =1$ universe.  In figures 3 and 4, we
sho
   w
the effect in the opposite case, namely assuming an $\Omega_0=1$ universe to
beg
   in
with, we examine both the angular diameter-redshift relations, and their
derivatives which result from several choices of $\alpha <0$.  As can be seen,
$\alpha =-.25$ results in a curve whose minima, and asymptotic slope are very
similar to the case of an $\Omega_0=.25, \Lambda =.75$ universe.
Alternatively,
a similar magnitude, but positive value for $\alpha$ would result in the latter
universe mimicking the former.

It is worth noting that the slopes of the curves, while similar at large
redshift, differ somewhat at small redshift, where most of the data will in
fact
be expected to be accumulated.  Could such a difference, if observed, be
observationally significant?  There is yet another instrinsic uncertainty which
suggests that unless one measures sources out to significantly beyond the any
observed minimum, comparison of the small redshift and large redshift slopes
may
be ambiguous.  Recall that we normalized the angular diameter so $D/R_0=1$.
Thi
   s
can only be done if we know the intrinsic value of $D$, or $D_0$ if $D$ varies.
However, our knowledge of this intrinsic magnitude suffers from our uncertainty
in the distance scale of universe, coming from our uncertainty in $H_0$, at the
factor of 2 level.  Changing $D$ by a factor of 2 will change the slope on
eithe
   r
side of the minimum by a similar factor, thus tilting the overall curve.
Unless
one has a sufficiently large lever arm on both ends, it is unlikely therefore
th
   at
one can overcome Hubble constant-based uncertainties in the slope of either
side
   .

Finally, is a value $|\alpha| =.25$ reasonable?  This suggests that at a
redshift of 2 the size of compact jets would have changed by about 30\%
compared
    to
their size in nearby galaxies.  This is not an extreme variation.  One could
imagine plausible mechanisms which could produce such an effect.  Radio jets
presumably result from accretion onto a black hole.  If the accretion rate
varie
   s
over cosmological time then this would impact on the energy
production which fuels the jets.  If, for example, heating by matter accreting
over time caused the material surrounding the hole to "puff up", this might
lower the accretion rate.  Alternatively, collapse of matter in the region of
the hole over time might increase the accretion rate.

These arguments are not meant to be fatal to this method.  There are merely
presented as caveats.  If a minimum {\it is} confirmed in the angular
diameter-redshift relation at $z \approx 1$ for radio jets then it will imply
on
   e of
three things: (a) the cosmic density parameter is greater than a certain
minimum
value which could exceed the value inferred from the dynamics of galaxies and
clusters, (b) if the universe is flat, the cosmological constant is smaller
than
the amount required to account for the difference between virial estimates for
clusters and galaxies and $\Omega =1$, or (c) evolutionary effects have the
same
form and magnitude as we have assumed here.  We expect the latter possibility
should be addressable by more detailed modelling.  If it can be invalidated,
then "universal lensing" causing a magnification of angular diameters of
objects
at cosmological redshifts can not only give strong evidence for a flat
universe,
but also can potentially rule out a cosmological constant dominated one.

\vskip .2in
\noindent{We thank B. Burke for informing us of the results of K. Kellerman,
and
K. Kellerman for kindly discussing his preliminary results with us. We also
than
   k
Martin White for checking the calculations of LMK for the redshifts
of minimum angular diameter as a function of $\Lambda$.}

\newpage

\parindent 0cm

\noindent {\bf \large References}

Baum, E., 1983, Phys. Lett. {\bf B133} 185

Coleman, S., 1984, Nucl. Phys. {\bf B310} 643

Fukugita, M.,Futamase,T. and Kasai,M. 1990, MNRAS {\bf 246} 24P

Fukugita, M. and Turner, E.L., 1992 MNRAS, in press

Hawking, S.W., 1984, Phys. Lett. {\bf B134} 403

Kellerman, K., proceedings of Conference on Sub Arc Second Radioastronomy,
Manchester, U.K. July 1992, to appear; also NRAO preprint to appear

Kochanek, C.S. 1992, Ap.J. {\bf 382} 46

Krauss, L.M. and White, M. 1992, Ap J. {\bf 394}, 385

Krauss, L.M., 1992, to appear in {\it Proceedings of Rencontres de Blois,
Partic
   le
Astrophysics}, (Editions Frontieres 1993)

Mao, S. 1991 Ap. J. {\bf 380} 9

Misner, C.W, Thorne, K.S. and Wheeler, J.A., 1973 {\it Gravitation} (Freeman,
Sa
   n
Francisco)

Pearson T.J. 1990, in {\it Parsec Scale Radio Jets} ed. by Zensus, J.A. and
Pearson, T.J. (Cambridge Univ. Press, Cambridge)

Smoot, G.F. et al, 1992, Ap. J., in press

Turner, E.L., 1990, Ap. J.Lett. {\bf 365} L43

Weinberg, S. 1972, {\it Gravitation and Cosmology} (Wiley, N.Y.)

\newpage

\begin{table}
\center
\begin{tabular}{||c|c|c||}           \hline
$\Omega_0$      &  Redshift of minimum $\delta$ & Asymptotic Slope (z=4) \\
\hli
   ne
0.0     & ---  &  -.062 \\ \hline
0.1     & 2.08  & .176  \\ \hline
0.2     & 1.76  & .254  \\ \hline
0.3     & 1.60  & .308  \\ \hline
0.4     & 1.50  & .353  \\ \hline
0.5     & 1.44  & .392  \\ \hline
0.6     & 1.38  & .426  \\ \hline
0.7     & 1.34  & .457  \\ \hline
0.8     & 1.31  & .486  \\ \hline
0.9     & 1.28  & .513  \\ \hline
1.0     & 1.25  & .538  \\ \hline
\end{tabular}

\caption{ Redshifts at Minimum and Asymptotic Slope of Angular Diameter
vs. Redshift curve for various values of $\Omega_0$ in a Flat Universe}
\end{table}

\vskip 1in

\noindent{\bf{Figure Captions}}
\vskip .1in

\noindent{Figure 1a: The angular diameter versus redshift for a unit fiducial
length perpendicular to the line of sight.  Cases shown include flat universes
with cosmological constant, and open universes with zero cosmological constant
}

\vskip .1in
\noindent{Figure 1b: The derivatives, with respect to redshift of the curves
sho
   wn
in figure 1a.  The redshifts of zero slope (i.e minima) occur where these
curves
intersect the heavy dashed curve.  Also important are the asymptotic values of
t
   he
slopes. }

\vskip .1in
\noindent{Figure 2a: Same as figure 1a for an $\Omega_0 =1$ flat universe with
evolutionary variation in fiducial refence probe going as
$D=D_0(1+z)^{\alpha}$,
    in
comparison to the cases $\Omega_0=0,1$ without such variation. }

\vskip .1in
\noindent{Figure 2b: same as figure 1b, for the cases described in figure 2a }

\end{document}